\documentclass[twocolumn,reprint,aps,prd,nofootinbib,floatfix,superscriptaddress]{revtex4-1}
 \usepackage[unicode]{hyperref} 
\usepackage[utf8]{inputenc}
\usepackage{amsmath}
\usepackage{amsfonts}
\usepackage{amssymb}
\usepackage[left=2cm,right=2cm,top=2cm,bottom=2cm]{geometry}
\usepackage{braket}
\usepackage{dsfont}
\usepackage{hyperref}

\usepackage{graphicx}
\usepackage{tikz}
\usetikzlibrary{arrows,decorations.pathmorphing,backgrounds,positioning,fit,petri,shapes}
\usepackage{lipsum}

\usepackage{multirow}
\begin{document}

\title{{\tt CombineHarvesterFlow}: Joint Probe Analysis Made Easy with Normalizing Flows}

\begin{abstract}
 We show how to efficiently sample the joint posterior of two non-covariant experiments with a large set of nuisance parameters. Specifically, we train an ensemble of normalizing flows to learn the posterior distribution of both experiments. Once trained, we can use the flows to reweight $\mathcal{O} (10^9)$ samples from both measurements to compute the joint posterior  in seconds -- saving up to $\mathcal{O}(1)$ ton of $\text{CO}_2$ per Monte Carlo run. Using this new technique we find joint constraints between the Dark Energy Survey $3 \times 2$ point measurement,  South Pole Telescope and Planck CMB lensing and a BOSS direct fit full shape analyses, for the first time. We find $\Omega_{\rm m} = 0.32^{+0.01}_{-0.01}$ and $S_8 = 0.79 ^ {+0.01}_ {-0.01}$. We release a public package called {\tt CombineHarvesterFlow}\footnote{\url{https://github.com/pltaylor16/CombineHarvesterFlow}} which performs these calculations.
\end{abstract}


\author{Peter L.~Taylor}
\email{taylor.4264@osu.edu}
\affiliation{Center for Cosmology and AstroParticle Physics (CCAPP), The Ohio State University, Columbus, OH
43210, USA}
\affiliation{Department of Physics, The Ohio State University, Columbus, OH 43210, USA}
\affiliation{Department of Astronomy, The Ohio State University, Columbus, OH 43210, USA}
\author{Andrei Cuceu}
\affiliation{Center for Cosmology and AstroParticle Physics (CCAPP), The Ohio State University, Columbus, OH
43210, USA}
\affiliation{NASA Einstein Fellow}
\author{Chun-Hao To}
\affiliation{Center for Cosmology and AstroParticle Physics (CCAPP), The Ohio State University, Columbus, OH
43210, USA}
\affiliation{Department of Physics, The Ohio State University, Columbus, OH 43210, USA}
\affiliation{Department of Astronomy, The Ohio State University, Columbus, OH 43210, USA}
\author{Erik A. Zaborowski}
\affiliation{Center for Cosmology and AstroParticle Physics (CCAPP), The Ohio State University, Columbus, OH
43210, USA}
\affiliation{Department of Physics, The Ohio State University, Columbus, OH 43210, USA}

\maketitle

\section{Introduction} \label{sec:intro}
\par Analyses of data from the Atacama Cosmology Telescope (ACT)~\cite{ACT:2023dou}, the Baryon Oscillation Spectroscopic Survey (BOSS)~\cite{eBOSS:2020yzd}, the Dark Energy Survey (DES)~\cite{DES:2021wwk}, the Dark Energy Spectroscopic Instrument (DESI)~\cite{DESI:2024mwx}, the extended Baryon Oscillation Spectroscopic Survey (eBOSS)~\cite{eBOSS:2020yzd}, Hyper Suprime-Cam (HSC)~\cite{More:2023knf}, the Kilo-degree Survey (KiDS)~\cite{Heymans:2020gsg}, Planck~\cite{Planck:2018vyg} and the South Pole Telescope (SPT)~\cite{SPT:2019fqo}, have provided some of the tightest cosmological constraints. Future measurements with Euclid~\cite{Euclid:2024yrr}, the Nancy Grace Roman Space Telescope\footnote{\url{https://roman.gsfc.nasa.gov/}}, the Rubin Observatory\footnote{\url{https://www.lsst.org/}}, the Simons Observatory\footnote{\url{https://simonsobservatory.org/}} and CMB-S4\footnote{\url{https://cmb-s4.org/}} will tighten these constraints even further. 
\par Meanwhile there has been a proliferation of cosmological likelihood packages to perform inference with these datasets (e.g. MontePython~\cite{Brinckmann:2018cvx}, CosmoMC~\cite{Lewis:2013hha}, COSMOLIKE~\cite{Krause:2016jvl}, desilike\footnote{\url{https://github.com/cosmodesi/desilike}}, CosmoSIS~\cite{Zuntz:2014csq}, Cobaya~\cite{Torrado:2020dgo}, firecrown\footnote{\url{https://github.com/LSSTDESC/firecrown}}, CLOE\footnote{Euclid Collaboration in prep.}). With each collaboration preferring different pipelines, performing joint analyses requires substantial effort and computational resources -- even when the analyses are statistically independent. It is this problem which we set out to solve.
\par Traditionally there have been three ways to sample from the joint posterior of two statistically independent measurements: 1) explicitly sampling the joint posterior, 2) importance sampling or 3) emulating the likelihood (or posterior) directly from the samples. All these approaches have significant drawbacks.
\par Explicitly sampling the likelihood requires the likelihood of each experiment to be integrated into a single unified likelihood package (e.g. CosmoSIS). This time consuming exercise may have to be repeated multiple times for each new experiment. The computational costs are also significant. A single DES Year 3 (DESY3) $3 \times 2$ point analysis takes up to 4 node days on a modern high-performance node with 128 cores~\cite{DES:2022ykc}.
\par Importance sampling offers a promising alternative. Suppose that we have access to samples from the posterior (chains) of one of the experiments, $\{ c_i\}$, over cosmological parameter space $\bold{c}$. Then to sample from the joint posterior we compute the likelihood of the second experiment at each point in the original chain $\{ \mathcal{L} (c_i) \}$ and then re-weight the original chain by these likelihoods. However, if both sets of experiments contain a large set of nuisance parameters, this option is no longer viable. This is because at every point in the first chain, there is a distribution of nuisance parameter values in the second experiment.
\par A third alternative is to directly emulate the likelihood (or posterior) using a neural network or Gaussian process emulator. In Appendices~\ref{sec:emu_1}-\ref{sec:emu_2}, we show that this approach fails in many circumstances. 
\par In this paper we advocate for a new approach using normalizing flows. Given two chains from two independent cosmological experiments our approach is capable of reweighting $\mathcal{O}(10 ^9)$ points from each experiment to compute the joint posterior in seconds. 
\par The structure of the paper is as follows. In Sec.~\ref{sec:formalism}, we present the formalism and describe our publicly released code in Sec.~\ref{sec:code}. In Sec.~\ref{sec:results}, we show our method works on a toy problem before presenting joint large scale structure parameter constraints from photometric, spectroscopic and CMB lensing data. Finally, in Sec.~\ref{sec:future} we discuss the future outlook for this method.

\section{Formalism} \label{sec:formalism}

\subsection{Joint Sampling Two Likelihoods} \label{sec:two likelihoods}
\par Let us suppose that we have performed two independent cosmological likelihood analyses and that we have samples from the posterior distributions, $P(\boldsymbol{c}, \bold{m_i} | \bold{d_i} )$,  for both experiments $i \in \{1,2\}$, assuming some model $\mathcal{M}$. Here, $\boldsymbol{c}$, are the parameters sampled in both chains, usually the cosmological parameters of interest, while $\bold{d_i}$ and $\bold{m_i}$ are the data vectors and nuisance parameters. In all that follows, we will assume that there is no covariance between the two data vectors and that the two experiments have completely different sets of nuisance parameters. We will also assume that the priors on the cosmologial and nuisance parameters are separable, that is $p(\bold{c}, \bold{m_i}) =  p(\bold{c})  p(\bold{m_i})$. These conditions are often met and we will apply them to a real world example in Sec.~\ref{sec:results}. For the time being we will also assume that the analyses were run with the same prior $p(\bold{c})$, but we relax this assumption in Sec.~\ref{sec:prior}.
\par To perform a joint analysis, we would like to find the joint posterior on $\bold{c}$, after marginalizing over both sets of nuisance parameters i.e.,
\begin{equation}
p(\bold{c}|\bold{d_1},\bold{d_2}) = \int \text{d} \bold{m_1} \text{d} \bold{m_2} \ p(\bold{c},\bold{m_1}, \bold{m_2} | \bold{d_1}, \bold{d_2}).
\end{equation}
Using the independence of the likelihood and Bayes' Theorem we find,
\begin{equation} \label{eq:result}
p(\bold{c}|\bold{d_1},\bold{d_2}) \propto \frac{p(\bold{c}| \bold{d_1})p(\bold{c}|\bold{d_2})}{p(\bold{c})},
\end{equation}
where, $p(\bold{c}| \bold{d_i})$, is 
\begin{equation}  \label{eq:def}
p(\bold{c}| \bold{d_i}) = \int \text{d} \bold{m_i} \ p(\bold{c},\bold{m_i} | \bold{d_i}).
\end{equation}
The factor $p(\bold{c})$ appears in the denominator of Eqn.~\ref{eq:result} to avoid double counting the prior. If the prior is uninformative, the denominator can be ignored. 
\par {\bf The key idea in this paper is that we can fit normalizing flows to estimate the conditional density, $p(\bold{c}| \bold{d_1})$, from the chains of the first experiment. Then to sample the joint posterior, we weight the samples of the second posterior by our estimate by our estimate $\widehat p(\bold{c}| \bold{d_1})$.} If the prior is informative we must also apply a weight $1/p(\bold{c})$, to all points, so that the final weight is $\widehat p(\bold{c}| \bold{d_1}) / p(\bold{c})$.
\par Alternatively, we could have trained the flows to approximate, $p(\bold{c}| \bold{d_2})$, and used this estimate to weight the samples of the first posterior. {\bf An essential cross-check is to ensure that we get the same joint posterior in both cases.} This will be discussed further in Sec.~\ref{sec:results}.
\par To combine $N$ analyses, we first combine the first two using Eqn.~\ref{eq:result} and then iteratively add the next $N-2$ analyses one at a time.

\subsection{Estimating the Marginal Posterior with Normalizing Flows}
Given samples from a target distribution $p_\bold{t}(\bold{x})$, one can train a normalizing flow to learn an invertible mapping, $F$, from a multivariate normal distribution centered at the origin with identity covariance, to the target. Once the mapping is learned, one can estimate the likelihood of each point in the target from the Jacobian: $p_\bold{t}(\bold{x}) = \mathcal{G}(F^{-1}(\bold{x})) |\frac{\partial F^{-1}}{\partial x}|$, where $\mathcal{G}$ is the probability density of the unit normal. We use this feature of normalizing flows to estimate the marginal posterior, $p(\bold{c}| \bold{d_i})$, from the chains as described in the previous section.
\par In all that follows, we use {\tt flowjax}\footnote{\url{https://github.com/danielward27/flowjax}}~\cite{ward2023flowjax} a publicly available package written in {\tt Jax} to train the normalizing flows using a masked autoregressive flow  (MAFs)~\cite{kingma2016improved, papamakarios2017masked} architecture. We use the default MAF settings in {\tt flowjax} throughout. We construct the invertible transform using a rational quadratic spline with 8 knots on the interval $[-4,4]$, using the built-in {\tt RationalQuadraticSpline} class in {\tt flowjax}. In all that follows the learning rate is set to $10^{-3}$.

\subsection{A Flow Ensemble Model}
\par To improve the accuracy of the density estimate, we train an ensemble of $N$ flows with the same architecture. Each flow is given equal weight so that when we compute the log-probability of a point $\bold{x}$, our estimate, $\ln \widehat p(\bold{x})$, is taken to be the average estimate over all the flows. We find that  results typically convergence for $N \gtrsim 5$.

\subsection{Using Outputs from Nested Samplers}
Nested samplers such as PolyChord~\cite{Handley:2015fda}, nautilus~\cite{Lange:2023ydq} and MultiNest~\cite{Feroz:2008xx} return a set of samples and corresponding weights. Since {\tt flowjax} does not take weights as input, we use a custom loss function in {\tt CombineHarvesterFlow} which uses the weighted average of the log-probabilities over the training set, rather than the average, used in {\tt flowjax}.

\subsection{Change of Variable} \label{sec:prior}
\par So far we have assumed that both sets of chains have the same priors. Let us relax this assumption and suppose the two chains have the priors, $p_1(\bold{c_1})$, and $p_2(\bold{c_2})$, and we have a target prior, $p_t(\bold{c_t})$, which we would like to use for the joint analysis. Here, the priors and target may not be expressed in the same parameterization but we will assume there is a mapping, $T: \bold{c_i} \rightarrow \bold{c_t}$, between $\bold{c_t}$ and $\bold{c_i}$, for each $i$. We will also assume that the support of $p(\bold{c_t})$ is a subset of the intersection of the supports of $p_1(\bold{c_1})$ and $p_2(\bold{c_2})$.
\par Then to map the posterior sample from $\bold{c_i}$ to $\bold{c_t}$, we weight the chains~\cite{Taylor:2017ipx} to account for both the change in prior and the Jacobian. Hence, if we write $c_{ij}$ for the $j$th point in chain $i$ and we assume it has an initial weight $w_I(c_{ij})$, then the final weight in the target parameterization is 
\begin{equation}
w_f\Big(T(c_{ij})\Big) = \frac{p_t\Big(T(c_{ij}) \Big)}{p_i(c_{ij})} \Bigg| \frac{\partial \bold{c_t}} {\partial \bold{c_i}} \Bigg| w_I (c_{ij}).
\end{equation}
After one has mapped both posteriors to the same target prior, one must still divide the joint constraints by $p(\boldsymbol{c_t})$ (see Eqn.~\ref{eq:result}), to avoid double counting the prior in the joint constraints.

\section{Code} \label{sec:code}
We release {\tt CombineHarvesterFlow} as a pip-installable package. The package is also available at \url{https://github.com/pltaylor16/CombineHarvesterFlow}. The code has a simple two class structure. The {\tt harvest} class allows the user to train, save and load flows on both CPU and GPU, while a {\tt combine} class allows the user to draw samples from the joint marginal posterior using the approach described in Sec~\ref{sec:two likelihoods}. {\bf Given the large number of potential corner cases, the package does not account for the prior in the denominator of Eqn.~\ref{eq:result}, nor weights associated with a change of variable described in Sec~\ref{sec:prior}}. The user is responsible for weighting the points in the input and output chains appropriately. The examples presented in Sec.~\ref{sec:toy}-\ref{sec:fail} are made publicly available in a {\tt jupyter} notebook released with the package -- in addition to a notebook containing an example with an informative prior.

\section{Results} \label{sec:results}

\subsection{Validation on Toy Models} \label{sec:toy}
As a simple test case, we find the joint constraints from two 2-dimensional Gaussian posteriors using {\tt CombineHarvesterFlow}. To validate our estimate, we use the fact that the product of two Gaussians with means $\boldsymbol{\mu_1}$ and $\boldsymbol{\mu_2}$ and covariances $\boldsymbol{\Sigma_1}$ and $\boldsymbol{\Sigma_2}$ is also a Gaussian with covariance $\boldsymbol{\Sigma} = [(\boldsymbol{\Sigma_1}) ^{-1} + (\boldsymbol{\Sigma_2})^{-1} ] ^{-1}$ and mean $\boldsymbol{\mu} = \boldsymbol{\Sigma}[(\boldsymbol{\Sigma_1})^{-1} \boldsymbol{\mu_1} + (\boldsymbol{\Sigma_2})^{-1} \boldsymbol{\mu_2}]$.
In this example, we assume there are no nuisance parameters and that the prior is uninformative. Before rotation, the mean and covariances of the two Gaussian are given by: ${\boldsymbol{\mu_1}} = (0,0)$, ${\boldsymbol{\mu_2}} = (1,0)$, $\boldsymbol{\Sigma_1} = {\rm diag}(1,4)$ and $\boldsymbol{\Sigma_2} = {\rm diag}(3,1)$. Both Gaussians are then rotated by $45 ^ \circ$.
\par
\begin{figure}[!hbt]
\includegraphics[width = \linewidth]{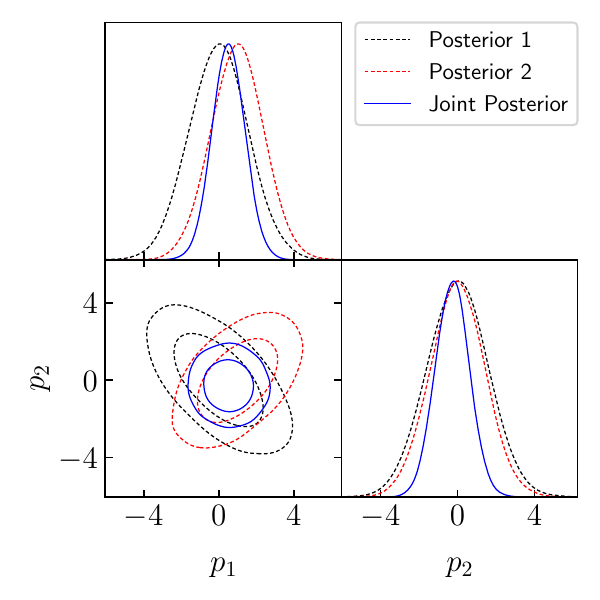}
\includegraphics[width = \linewidth]{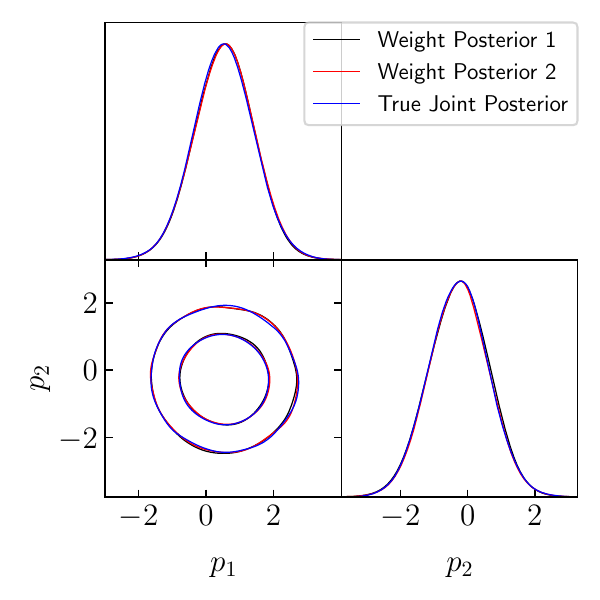}
\caption{{\bf Top:} The $68 \%$ and $95 \%$ confidence intervals for two initial Gaussian posteriors are indicated by the red and black dashed lines. The joint constraints which are known analytically are shown in solid blue. {\bf Bottom:} The joint constraints are again shown in blue, while the constraints found using {\tt CombineHarvesterFlow} from weighting the first posterior and the second posterior are shown in black and red respectively. In both cases, {\tt CombineHarvesterFlow} accurately recovers the joint posterior.}
\label{fig:toy_model_1}
\end{figure}

 The resulting $68 \%$ and $95 \%$ confidence regions on parameters $p_1$ and $p_2$ found by sampling 20000 points from these distributions are shown in the top panel of Fig.~\ref{fig:toy_model_1} by the dashed black and red contours respectively. The confidence regions of the joint posterior found by sampling 20000 points from the exact joint posterior is shown in solid blue.  
\par Using {\tt CombineHarvesterFlow}, we train an ensemble of 7 flows to learn the original posteriors. We then weight the samples in the first posterior by the flow weights of the second posterior -- and crucially as a cross-check, we weight the samples of the second posterior by the flow weights of the first posterior. The results are shown in the bottom panel of Fig.~\ref{fig:toy_model_1}. In both cases, we reproduce the true joint posteriors almost exactly.

\subsection{Validation of {\tt CombineHarvesterFlow} and Causes of Failure} \label{sec:fail}

\begin{figure*}[!hbt]
\includegraphics[width = 5.5cm]{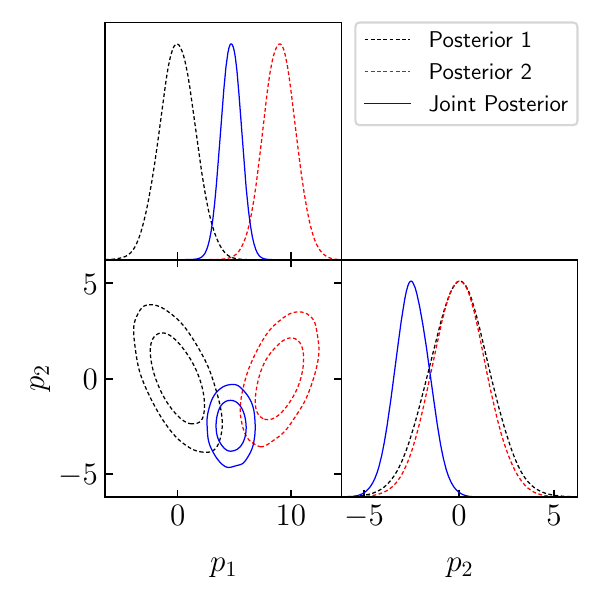}
\includegraphics[width = 5.5cm]{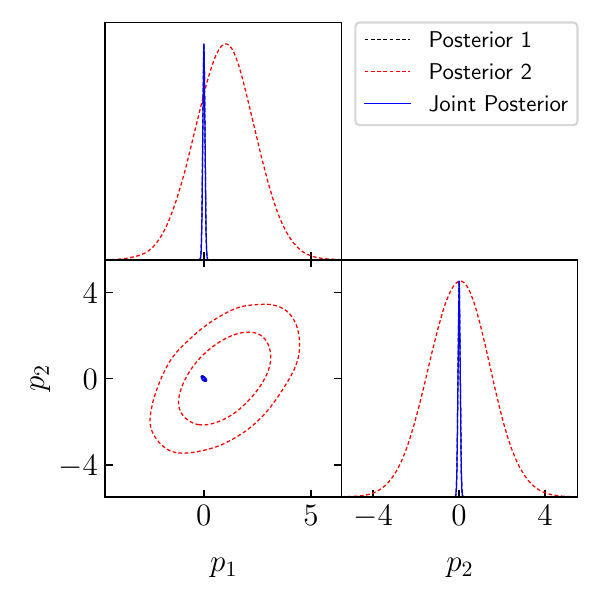}
\includegraphics[width = 5.5cm]{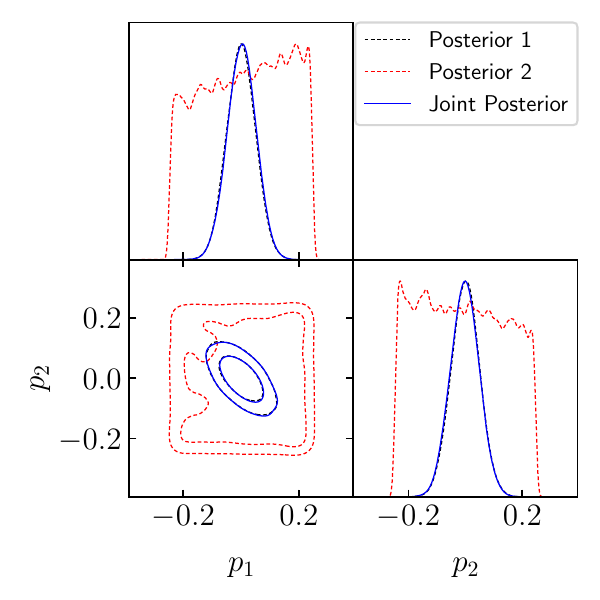}
\includegraphics[width = 5.5cm]{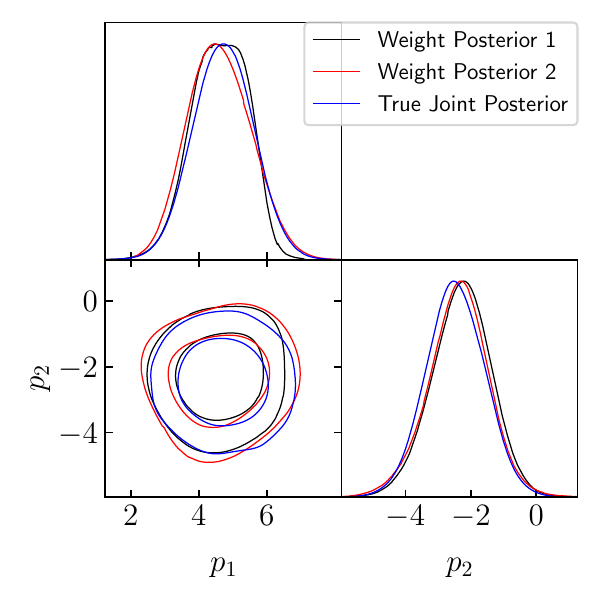}
\includegraphics[width = 5.5cm]{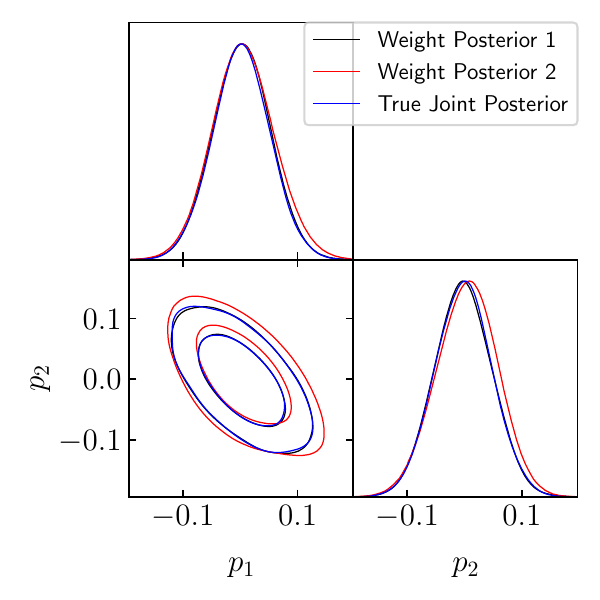}
\includegraphics[width = 5.5cm]{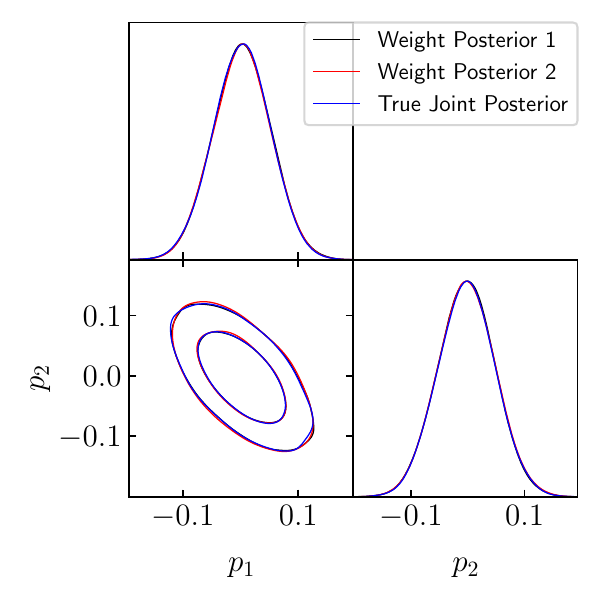}
\caption{{\tt CombineHarvesterFlow} can fail due to under sampling regions of high joint density ({\bf left} and {\bf right} panels), but can be fixed in certain circumstances by adjusting the prior {\bf right}. Failures are usually caused by tensions between the two experiments ({\bf left}), or when one experiment is much more constraining than the other ({\bf middle}). In former case one should not combine experiments as the tension likely points to systematics -- while in the later case, the tension can be overcome by enforcing a tighter prior on the less constraining experiment ({\bf right}). {\bf Top:} Dashed lines indicate the original $68 \%$ and $95 \%$ confidence intervals for two experiments, while solid blue indicates the true joint posteriors. {\bf Bottom:} Black and red contours are the {\tt CombineHarvesterFlow} estimates from weighting the first and second posteriors, while blue indicates the true joint constraints. In the case of tension (left) or undersampling (middle) the two {\tt CombineHarvesterFlow} joint posterior estimates do not match so the results should be rejected.}
\label{fig:toy_model_2}
\end{figure*}

In certain circumstances {\tt CombineHarvesterFlow} can fail to return accurate joint posteriors. In this Section, we discuss the most common causes and show that checking that the results derived from weighting the first posterior versus the second posterior is a robust way to ensure {\tt CombineHarvesterFlow} returns accurate constraints.
\par If two experiments are in severe tension, one should not attempt to find joint constraints, as the tension likely indicates the presence of systematics -- or that the underlying model assumption (e.g. $\Lambda$CDM) is wrong. However, if one does persist to find joint constraints, this is an instance where {\tt CombineHarvesterFlow} can fail. This is because the sampling density is so low in the regions where the joint posterior is large, that the normalizing flows do not accurately model the density in these regions.
\par To illustrate this point, we consider two Gaussian posteriors with means and covariances given by: ${\boldsymbol{\mu_1}} = (0,0)$, ${\boldsymbol{\mu_2}} = (9,0)$, $\boldsymbol{\Sigma_1} = {\rm diag}(1,4)$ and $\boldsymbol{\Sigma_2} = {\rm diag}(3,1)$, before rotating both Gaussians by $45 ^ \circ$. These posteriors are shown as dashed black and red lines in the top left of Fig.~\ref{fig:toy_model_2}. The joint constraints are shown in solid blue. In the bottom left corner of Fig.~\ref{fig:toy_model_2}, we zoom in on the joint constraints. The {\tt CombineHarvesterFlow} estimates from weighting the first and second posteriors are shown in black and red respectively, while the true joint constraints are shown in blue. In both cases {\tt CombineHarvesterFlow} fails to accurately reproduce the joint posteriors. The two {\tt CombineHarvesterFlow} constraints are in disagreement which suggests the {\tt CombineHarvesterFlow} results should be discarded.
\par {\tt CombineHarvesterFlow} can also fail when one experiment is much more constraining than the other. This is because the less constraining experiment undersamples the region where the joint density is large. 
\par To exemplify, we again consider two Gaussians. This time we take ${\boldsymbol{\mu_1}} = (0,0)$, ${\boldsymbol{\mu_2}} = (1,0)$, $\boldsymbol{\Sigma_1} = {\rm diag}(0.001,0.004)$ and $\boldsymbol{\Sigma_2} = {\rm diag}(3,1)$. The original posteriors and joint constraints are shown in the top middle panel of Fig.~\ref{fig:toy_model_2} while the true joint posterior and {\tt CombineHarvesterFlow} estimates are shown in the bottom middle panel. We see that weighting the more tightly constraining posteriors does actually return an accurate estimate of the joint posterior, while weighting the less constraining posterior results in slightly different constraints. Even though weighting the tighter posteriors accurately determined the joint posterior in this instance -- it is not clear that this will be true in general, so we always recommend discarding the results when the two {\tt CombineHarvesterFlow} estimates are not the same.
\par In the case where one set of posteriors is much tighter than the other, there is likely little to be gained from combining the results. Nevertheless, we can still obtain joint constraints if we first place a more restrictive prior on the less constraining experiment and rerun the chains to increase the sampling density in the relevant region of parameter space. This still may be favorable to running joint chains if the two likelihood pipelines are not merged.
\par To highlight this point, we take the same initial posteriors as in the previous example, but this time enforce a flat prior so that $p_i \in [-0.2,0.2]$. We show the initial posteriors in the top right of Fig.~\ref{fig:toy_model_2} and the joint posterior and the {\tt CombineHarvesterFlow} estimates are shown in the bottom right panel. In this case the two {\tt CombineHarvesterFlow} estimates and the true joint posteriors agree exactly. 
\par Even in the extreme examples presented in this section, {\tt CombineHarvesterFlow} extracts reasonably accurate posteriors -- while the results from the two different weighting strategies give a rough estimate on the uncertainties. In many circumstances (e.g. non-publishable testing and validation runs) this accuracy is more than sufficient, so we expect {\tt CombineHaresterFlow} to dramatically accelerate workflows even when it fails to produce extremely accurate joint posteriors. We stress that these errors are typically much smaller than those expected from model mis-specification. In the future, we recommend that users specify a success criterion (e.g. fractional error on the error for parameters of interest) and add the differences from the {\tt CombineHaresterFlow} cross-check to the total error budget.

\subsection{DESY3 $\times$ CMB Lensing + Boss Full Shape Direct Fit}
Due to the different coverage in radial modes, the covariance between projected statistics and spectroscopic full shape statistics is negligible~\cite{Taylor:2022rgy}. Hence, we use {\tt CombineHarvesterFlow} to derive joint constraints between the DESY3 $\times$ CMB lensing $6 \times 2$ point posterior\footnote{publicly available at \url{https://des.ncsa.illinois.edu/releases/y3a2}}~\cite{DES:2022urg} and the BOSS full shape direct fit posteriors found in~\cite{Philcox:2021kcw}\footnote{We use the chains which exclude the bispectrum information}. To our knowledge this is the first time that this combination of probes has been combined into a single analysis. We pay particular attention to $S_8$ where tensions between CMB and large scale structure measurements have previously been reported (see e.g. Fig. 12 in~\cite{Lange:2023khv}).
\par We use {\tt CombineHarvesterFlow}\footnote{We trained an ensemble of 50 flows and checked that our results are converged.} to find joint constraints in $\{ \Omega_{\rm m}, h_0, \omega_{\rm b}, n_{\rm s}, S_8 \}$. We take flat uninformative priors on $h_0$ and $A_{\rm s}$.\footnote{We note however, that our choice of $n_s$ prior can have a substantial impact on the inferred value of $h_0$ in CMB lensing~\cite{Kable:2023bsg}.} Since the DESY3 $\times$ CMB lensing constraints ran up against the boundaries of the flat prior on $n_{\rm s}$, we impose a conservative Gaussian prior, $n_{\rm s} \sim \mathcal{N}(0.96, 0.02)$, by retroactively weighting the chains after running {\tt CombineHarvesterFlow} to avoid boundary effects. The DES chains were parameterised in $ \{ \Omega_{\rm m}, h_0, \Omega_{\rm b}, n_{\rm s}, A_{\rm s} \}$ space while the BOSS chains were run in  $ \{ \omega_{\rm c}, h_0, \omega_{\rm b}, n_{\rm s}, \ln (10^{10} A_{\rm s}) \}$. Hence, the contribution of the Jacobians cancel when going to $\{ \Omega_{\rm m}, h_0, \omega_{\rm b}, n_{\rm s}, A_{\rm s} \}$-space, when we introduce a weighting factor of $A_{\rm s}$ to the chains. Both chains, store the $\sigma_8$ values at all sampled points, so we train the flows and present our final results in $S_8$, rather than $A_{\rm s}$. Finally, we take a Gaussian prior on $\omega_{\rm b} \sim N (0.02268, 0.00036)$ as in the BOSS chains. Since the DES prior is uninformative, this is achieved simply by not dividing out by the $\omega_{\rm b}$ prior in Eqn.~\ref{eq:result}.

\begin{figure}[!hbt]
\includegraphics[width = \linewidth]{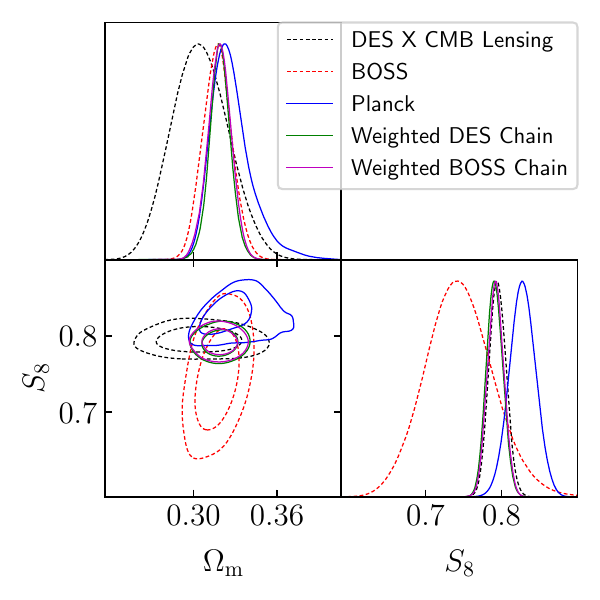}
\caption{Joint DESY3 $\times$ CMB lensing and BOSS full shape direct fit $68 \%$ and $95 \%$ confidence intervals. When running {\tt CombineHarvesterFlow} we find good agreement between weighting the BOSS and DES chains. The joint $S_8$ constraints are consistent with Planck (blue) at the $1.3 \sigma$ level.}
\label{fig:joint}
\end{figure}

\par The joint constraints in the $S_8 - \Omega_{\rm m}$ plane\footnote{The full joint constraints are shown and discussed in Appendix~\ref{sec:full constraints}.} are shown in Fig.~\ref{fig:joint}. Crucially, we find good agreement between weighting the DES $6 \times 2$ point chains and the BOSS chains. 
We find that the two {\tt CombineHarvesterFlow} constraints are consistent and $\Omega_{\rm m} = 0.32^{+0.01}_{-0.01}$ and $S_8 = 0.79 ^ {+0.01}_ {-0.01}$. For comparison, reweighing the Planck chains\footnote{We start with the the Planck chains available at {\url https://des.ncsa.illinois.edu/releases/y3a2} to match the DES prior.} to have the same prior described above, we find $\Omega_{\rm m} = 0.32^{+0.02}_{-0.02}$ and $S_8 = 0.83 ^ {+0.02}_ {-0.02}$. Adding the $S_8$ errors in quadrature~\cite{DES:2020hen} the joint BOSS and DES $6 \times 2$ point constraints are consistent with Planck at the $1.3 \sigma$ level.

\section{Future Outlook} \label{sec:future}

\subsection{Environmental and Cost Impact}
As noted in~\cite{To:2022ubu} (hereafter To22), the environmental impact of single Monte Carlo analysis are $\mathcal{O} (\$100)$ in energy costs and emit $\mathcal{O} (1)$ ton of $\text{CO}_2$. This is roughly equivalent to the emission of a transatlantic flight. Over the next decade, collaborations may collectively run as many as $ \mathcal{O} (10^2) - \mathcal{O} (10^4)$ joint analyses, in which case our approach could save up to $\mathcal{O} (\$ 10,000) - \mathcal{O} (\$ 1,000,000)$ in energy costs and reduce emissions by $\mathcal{O}(10^2) - \mathcal{O}(10^4)$ tons of $\text{CO}_2$.

\subsection{Methodological Extensions}
In this work we have focused on combing constraints from two-point statistics but we suspect that due to the different radial coverage in Fourier space that the covariance between projected and redshift-space space analyses is negligible in general. This opens the possibility of performing joint non-Gaussian constraints derived from non-Gaussian estimators, simulation-based inference or field-level Hamiltonian Monte-Carlo (e.g.~\cite{Nguyen:2024yth, Porqueres:2021clw,Hou:2024blc,Valogiannis:2022xwu,Hahn:2023udg,Cheng:2020qbx,DES:2024jgw,DES:2024xij}). 
\par Several papers have investigated using normalizing flows to estimate the Bayesian evidence directly from MCMC chains~\cite{Srinivasan:2024uax,Polanska:2024arc}. In a follow up work we will investigate computing the evidence of the joint posteriors. This will require training the flows on the full posteriors, $p(\bold{c}, \bold{m_i})$.

\section{Conclusion} \label{sec:conclusion}
In this paper we have shown how to jointly sample from two independent experiments with a large set of nuisance parameters using only the chains from the two experiments. In particular, training normalizing flows to learn the posterior, after marginalizing over the nuisance parameters, and then cross-weighting the chains by the probability density effectively recovers the joint posterior when there is no covariance. We release a public package called {\tt CombineHarvesterFlow} which performs these calculations.
\par {\tt CombineHarvesterFlow} enables joint probe parameter estimation without tedious pipeline integration. One can even sample from joint likelihoods when the likelihood pipeline for one of the datasets is not publicly available. \par {\tt CombineHarvesterFlow} is fast. We have found that training time for each flow can vary depending on the hardware and size of parameters space -- but training typically takes less than a minute on a CPU. Once trained, the flows enable virtually instantaneous sampling of the joint posterior.  
\par We estimate that compared to running joint likelihood analyses our approach will reduce emissions by between $\mathcal{O} (10^2) - \mathcal{O} (10^4)$ tons of $\text{CO}_2$ and save between  $\mathcal{O} (\$ 10,000) - \mathcal{O} (\$ 1,000,000)$ in energy costs.
\par This speed and efficiency opens up new possibilities for validation tests inside large collaborations. For example, when different parts of the data vector are not covariant (e.g. different tracers in DESI or the photometric and spectroscopic analyses in Euclid), one can run independent MCMC runs for each part of the data vector and then combine different combinations --determining the cosmological impact of each piece -- at virtually no additional computational cost. 
\par Our approach can fail when two experiments are in tension or when one measurement is significantly more constraining than the other. In the former case, one should not combine the measurements as the tension likely signals the presence of unmitigated systematics, while in the later case there is little to be gained from combining the measurements.
\par {\tt CombineHarvesterFlow} helps the user detect spurious results. In particular, weighting the first chain by the flows trained on the second chain, and vice versa, should yield consistent constraints. After submission, we became aware of several papers which proposed similar techniques to the one proposed in this work~\cite{Bevins:2022wsc, Bevins:2022qsc}. Subsequent to this work~\cite{Mootoovaloo:2024sao} also proposed a similar method. However, the ability to detect spurious results with the cross-check is a key feature that sets {\tt CombineHarvesterFlow} apart.
\par We have also shown that when {\tt CombineHarvesterFlow} fails due to one posterior being much tighter than the other, readjusting the original priors and resampling can resolve the issue. For this reason, {\bf we recommend that there is coordination between collaborations to choose priors that ensure chains can be reused without costly resampling.}
\par Since the covariance between projected angular statistics and redshift space multipoles is negligible, we used {\tt CombineHarvesterFlow} to combine DESY3 $\times$ CMB lensing with BOSS full shape direct fit measurements. To our knowledge this is the first time  a joint analysis of these datasets have been performed. We find $S_8 = 0.79 ^ {+0.01}_ {-0.01}$ which is consistent with Planck at the $1.3 \sigma$ level. This example showcases the efficacy of {\tt CombineHarvesterFlow} -- a tool which will render many currently tedious joint sampling problems trivially implementable, fast and computationally inexpensive.

\section{Acknowledgements}
 PLT is supported in part by NASA ROSES 21-ATP21-0050. This work received support from the U.S. Department of Energy under contract number DE-SC0011726. AC acknowledges support provided by NASA through the NASA Hubble Fellowship grant HST-HF2-51526.001-A awarded by the Space Telescope Science Institute, which is operated by the Association of Universities for Research in Astronomy, Incorporated, under NASA contract NAS5-26555. The authors thank Mikhail Ivanov for providing the BOSS full shape chains used in this work and the anonymous referee who's comments improved the manuscript.  We acknowledge use of the open source software~\cite{jax2018github, Hinton2016, Lewis:2019xzd, 2020SciPy-NMeth, harris2020array, Hunter:2007}
 
 \bibliographystyle{apsrev4-1.bst}
\bibliography{bibtex.bib}

\appendix

\section{Failure of Neural Network Likelihood Emulation} \label{sec:emu_1}

\begin{figure}[!hbt]
\includegraphics[width = \linewidth]{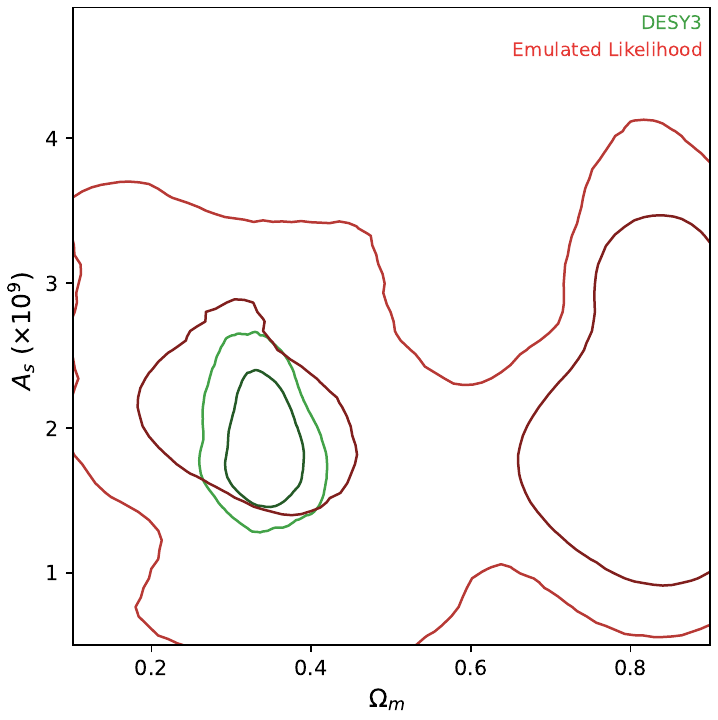}
\caption{Emulating the likelihood directly from the chains fails catastrophically because the emulator is completely unconstrained over most of the prior volume. {\bf Green:} Fiducial DESY3 $3 \times 2$ point analysis. {\bf Red:} Reanalysis using an emulated likelihood.}
\label{fig:emulator_1}
\end{figure}

Emulating the data vector over parameter space using neural networks is a popular technique to accelerate parameter inference (see e.g.~\cite{To:2022ubu,Saraivanov:2024soy}). Since we wish to compute posteriors from the output chains, we do not have access to the data vectors, but we can emulate the likelihood if the chains also store the likelihood at the sampled points. Emulating both likelihoods could in principle dramatically speed up joint parameter inference.
\par However, we find that direct emulation of the likelihood, rather than data-vector can fail catastrophically. This is because most of the prior volume is not sampled by the chains leaving the emulated likelihood completely unconstrained in these regions.\footnote{Emulating the data-vector does not suffer from this issue because in unconstrained regions the $\chi^2$ is most likely large and hence rejected by the sampler.} 
\par  To illustrate this, we trained a likelihood emulator using the publicly available DESY3 $3 \times 2$ point chains. Using this emulator we reran the DES chains, sampling with Emcee~\cite{foreman2013emcee}. The results in the $\Omega_{\rm m} -A_{\rm s}$ plane, after marginalizing over the other parameters, are shown in Fig.~\ref{fig:emulator_1}. Here it is clear the emulated likelihood analysis has failed. It may be possible to solve this problem by (e.g. forcing the posterior to zero away from the samples), but this could require substantial fine-tuning to the problem at hand.

\section{Failure of Gaussian Process Likelihood Emulation} \label{sec:emu_2}
\begin{figure}[!hbt]
\includegraphics[width = \linewidth]{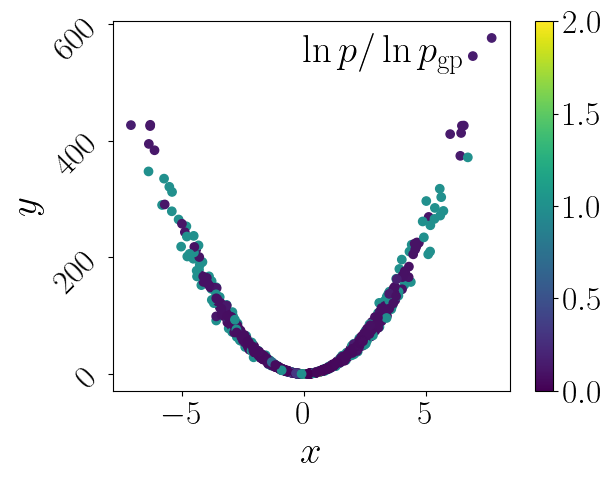}
\caption{A sample of 1000 points from a highly non-Gaussian likelihood described in Appendix~\ref{sec:emu_2}. Colors represent the ratio of the true likelihood to the GP-emulated likelihood. For many of the points, the ratio is far away from unity indicating that the emulator has failed dramatically. This is due to the GP sampling scheme which fails for non-Gaussian posteriors.}
\label{fig:emulator_2}
\end{figure}

The authors of~\cite{McClintock:2019ijs}, proposed direct emulation of the likelihood using Gaussian processes (GP) and released a GP emulation package, {\tt AReconstructionTool}\footnote{\url{https://github.com/tmcclintock/AReconstructionTool}}. Unlike neural network emulators, the GP-emulated likelihood is well constrained outside the immediate vicinity of the training set. However, GP emulators do not scale with large training sets which requires one to subsample the chains~\cite{McClintock:2019ijs}. We find that the sampling scheme available {\tt AReconstructionTool}, which is optimized for Gaussian posteriors such as those found in Planck, does not generalize well to non-Gaussian posteriors.
\par As an example, we take the Gaussian with mean $(\mu_x,\mu_y) = (0,10)$ and covariance, $\boldsymbol{\Sigma} =  {\rm diag}(5,1)$ and make the transform $y \rightarrow y'=  x^2 y$. We compute the log-likelihood, which is analytically tractable, at 1000 points randomly sampled from the likelihood. At the same points we estimate the ratio of the true likelihood to the emulated likelihood. The results are shown in Fig.~\ref{fig:emulator_2}. We find that for many of these points, the ratio is far from unity indicated the GP-emulator has failed dramatically.

\section{Full Constraints} \label{sec:full constraints}

\begin{figure*}[!hbt]
    \begin{centering}
        \includegraphics[{width=1.0\textwidth}]{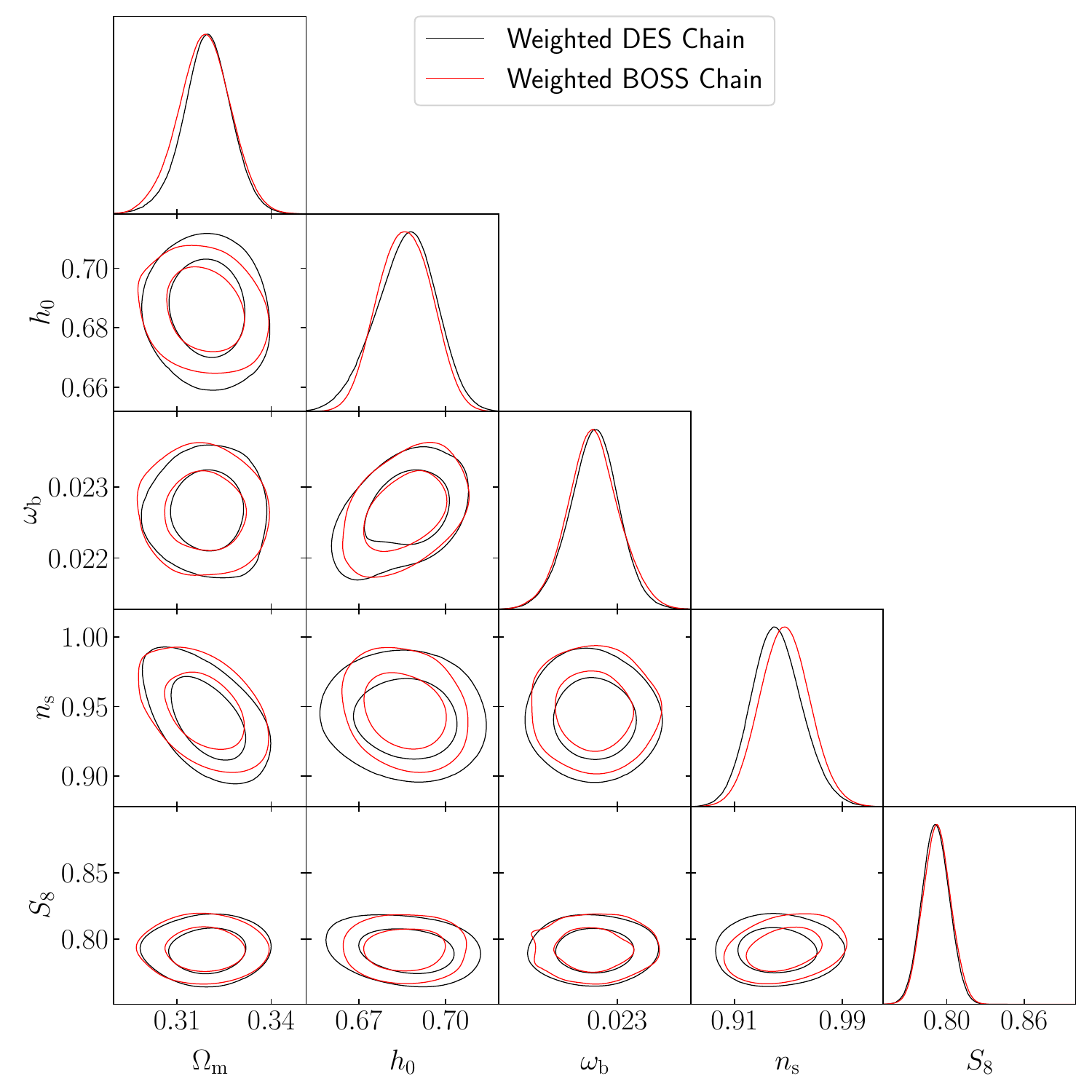}

        \caption{{\tt CombineHarvesterFlow} joint constraints from DESY3 $\times$ CMB lensing and BOSS full shape analyses. Different weighting strategies result in consistent constraints in the parameters of interest, $S_8$ and $\Omega_{\rm m}$, but the other parameters are not completely consistent. This is likely due to undersampling, as the DES posteriors are much broader in $h_0$ and $\omega_b$.}
    
    \label{fig:full}
    \end{centering}
    \end{figure*}
        
We plot the full joint constraints from the DESY3 $\times$ CMB lensing and BOSS full shape analyses in Fig.~\ref{fig:full}. Although we find very consistent results when weighting the DES chains compared to weighting the BOSS chain in the main parameters of interest, $S_8$ and $\Omega_{\rm m}$, we find small discrepancies in the other parameters.
\par These differences are almost certainly caused by the DES chains undersampling the regions where the joint posterior is high (see Sec.~\ref{sec:fail}). In particular, to calibrate the BAO scale, the BOSS chains prior on $\omega_b$ is substantially tighter than the initial prior in DES chains and DES alone places virtually no constrain on $h_0$. This reinforces the need for collaborations to agree on priors which enable easy parameter combination a posteriori with {\tt CombineHarvesterFlow}.

\end{document}